# Critical Parameters, Thermodynamic Functions, and Shock Hugoniot of Aluminum Fluid at High Energy Density

Mofreh R. Zaghloul

Department of Physics, College of Sciences, United Arab Emirates University,

P.O.B. 15551, Al-Ain, UAE.

**ABSTRACT**

We present estimates of the critical properties, thermodynamic functions, and principal shock Hugoniot of hot dense aluminum fluid as predicted from a chemical model for the equation-of-state of hot dense, partially ionized and partially degenerate plasma. The essential features of strongly coupled plasma of metal vapors, such as multiple ionization, Coulomb interactions among charged particles, partial degeneracy, and intensive short range hard core repulsion are taken into consideration. Internal partition functions of neutral, excited, and multiply ionized species are carefully evaluated in a statistical-mechanically consistent way. Results predicted from the present model are presented, analyzed and compared with available experimental measurements and other theoretical predictions in the literature.

## 1. INTRODUCTION

Investigating the equation-of-state and thermodynamic properties of strongly coupled, partially ionized and partially degenerate plasmas is getting a growing interest in the literature. These plasmas are encountered in astrophysics and in many high-energy-density laboratory experiments such as laser interaction with metal targets [1], exploding wires [2], and wire array Z-pinch [3]. Aluminum plasma is an example of such strongly coupled systems which are generated over a wide range of densities and temperatures in many of such high-energy-density applications stimulating a continuous need to study and investigate the properties of these plasmas. The need for accurate prediction of the properties of aluminum (Al) plasma over a wide region of the density-temperature ($\rho$-$T$) phase space is particularly augmented by the growth of interest in new launch technologies [4,5], pulsed power machines designed to produce inertial confinement fusion plasmas [3,6] and for being a common component in shock experiments and space type vehicles. Such predictions for aluminum plasmas necessitate information about the population densities of different plasma species (neutral, excited, ionized, etc.) which require careful consideration of the ionization equilibrium problem. Special interest has to be devoted to the ultrahigh pressure phenomenon of pressure ionization for its crucial rule in understanding and explaining relevant experimental results.

A fundamental associated problem is the determination of the parameters of the critical point (critical temperature $T_c$, density $\rho_c$, and pressure $P_c$) of the material. Information about the critical properties of the material is required for understanding the fundamental mechanisms for many physical phenomena and processes such as phase-explosion encountered in many applications [7]. In view of the difficulty of measuring the critical properties of Al because of the pertinent extreme conditions, several theoretical estimates of these properties have been introduced in the literature. These estimates scatter over a relatively wide range of values as it can be seen from the representative collection given in Table 1.

**Table 1** A representative collection of estimated values of the critical properties of Al in the literature

| Reference | $T_c$ [K] | $P_c$ [MPa] | $\rho_c$ [kg m$^{-3}$] |
|---|---|---|---|
| Young and Alder 1971 [8] | 7151 | 545.8 | 690 |
| Ohse and Tippelskirch 1977 [9] | 4744 – 8550 | 171.7-545.8 | 400-690 |
| Martunjuk 1983 [10] | 5410 | 83 | 490 |
| Fortov et al. 1990 [11] | 8000 | 447 | 640 |
| Gathers 1994 [12] | 5726.5 | 182.02 | 423.56 |
| Hess 1998 [13] | 8944 | 472.6 | 430 |
| Likal'ter 1985, 2000, 2002 [14-16] | 9300<br>8000—8860<br>8860 | 570<br>468-447<br>312 | <br><br>280 |
| Basin 2002 [17] | 5445 | 192 | 551 |
| Singh et al 2006 [18] | 8472 | 509.4 | 785 |
| Bhatt et al 2006 [19] | 6300 | 88.4 | 707 |
| Ray et al 2006 [20] | 5700 | 187 | 320 |
| Lomonosov 2007 [21] | 6250 | 197 | 703 |
| Gordeev et al 2008 [22] | 7917 | 467 | 660 |
| Morel et al 2009 [23] | 6700±800 | N/A | N/A |
| Faussurier et al. 2009 [24] | 7963 | 350 | 440 |
| Belashchenk et al 2011 [25] | 7050 | 325 ± 20 | 675 ± 34 |
| Karabutov et al 2012 [26] | 7963 | 350 | - |
| Mishra and Chaturvedi 2012 [27] | 8387 | 445 | 385 |

An important means for benchmarking and validating theoretically generated equations-of-state is through comparison with the experimental measurements of the shock Hugoniot. The shock Hugoniot is the locus of all final states characterized by ($\rho, P, T$) that can be reached behind each shock for a sequence of different-strength shocks from a given initial state. Measuring the shock Hugoniot from shock wave experiments is a well-established technique that has been applied to a wide range of materials including aluminum, generating a substantial data set for Al that can be used for benchmarking and validation of the present model predictions.

## 2. CHEMICAL MODEL AND FREE ENERGY MINIMIZATION

Aluminum vapor is, in general, a complex mixture of neutral and charged clusters (more than one atom/ion) at different excitation states, neutral atoms ($Al_0$) at different excitation states, and multiply charged ions ($Al_1$, $Al_2$, $Al_3$,....,$Al_{12}$) at different excitation states in addition to bare nuclei ($Al_{13}$) and free electrons ($e$). In the chemical model (see for example, [28-31]), composite structured particles such as clusters, neutral atoms or ions are treated as elementary members of the thermodynamic ensemble together with bare nuclei and free electrons. Using some simplifying assumptions, it is possible, using statistical mechanics, to express the bulk-state properties of such an assembly in terms of relevant properties of individual particles. This is commonly achieved through the construction of an additive or separable free energy function where interactions or couplings among different species are usually accounted for through a configurational free energy component and a scheme for truncating the internal partition functions. For the range of temperatures considered herein (from marginally below the critical point to very high temperatures) it is unlikely that neutral and charged clusters exist in high concentrations that significantly affect the equation-of-state. Therefore, the formation and existence of neutral and charged clusters may be tolerably neglected and the assembly can be effectively considered as a mixture of different excitation states of atomic aluminum, $Al_0$, and positively charged ions ($Al_1$, $Al_2$,....., $Al_m$,....., $Al_{12}$), in addition to bare nuclei, $Al_{13}$, and free electrons, $e$. The sum of the separable components of the free energy can, therefore, be written as;

$$F = \sum_{m=0,\ldots,13} F_{Al_m, ideal} + F_{e, ideal} + F_{corr} + F_{zero-of-energy}, \qquad (1)$$

The first two components in the right hand side of Eq. (1) are the classical ideal free energy of heavy particles (atom/ions) and the classical ideal free energy of the free electrons while $F_{corr}$ is a term that takes into account possible quantum mechanical (degeneracy) and configurational effects including van der Waals' corrections while the $F_{zero-of-energy}$ term takes into account and corrects for the fact that the free energy components in Eq. (1) must all be calculated using energies referred to the same reference or "*zero-of-energy*".

At very high temperatures (>1,000,000 K) the contribution of the photon gas to the thermodynamics of the resulting plasma system may become important and one needs to add to the right hand side of Eq. (1) the free energy of the photon gas which can be expressed, for a blackbody radiation, as [32];

$$F_{bb\_rad} = -(4\sigma/3c)\, V\, T^4 \qquad (2)$$

where $\sigma$ is the Stefan-Boltzmann constant, $T$ is the absolute temperature, and $c$ is the speed of light.

Among ions of different ion multiplicities and free electrons there exit a set of ionization, and recombination processes. Assuming no nuclear reactions can take place, the minimization of the free energy function, $F$, for the ionization/recombination equilibrium reaction $Al_m \leftrightarrow Al_{(m+1)} + e$ requires that

$$\frac{\partial F}{\partial N_{Al_m}} - \frac{\partial F}{\partial N_{Al_{(m+1)}}} - \frac{\partial F}{\partial N_e} = 0, \qquad for\ \ m = 0,1,2,\ldots,Z_{nuc}-1 \qquad (3)$$

In Eq. (3), $N_{Al_m}$ and $N_e$ are the occupation numbers of Al ions of multiplicity $m$ ($m=0$ corresponds to neutral Al atoms) and free electrons, respectively while $Z_{nuc}=13$ is the atomic number. Since the present study is restricted to aluminum, we will drop the chemical symbol, *Al*, and simply use the multiplicity, *m*, to refer to *Al_m* in the following discussion.

Solving the system of equations in (3) subject to conservation of electric charge and conservation of nuclei gives the required equilibrium composition. Upon determining the equilibrium composition one can proceed to calculate the set of thermodynamic properties using standard thermodynamic relations. It is worth mentioning that in liquids at sufficiently high temperatures (slightly below the critical temperature) atoms move freely through the body in a way similar to their random motion in the gas phase. One can, therefore, and to a good approximation, use the free energy formulation of the vapor phase to represent the liquid phase in the neighborhood of the critical point.

## 3. IDEAL CLASSICAL FREE ENERGY COMPONENTS

The classical-ideal free energy of atomic aluminum or aluminum ions of different multiplicities is given by the classical expression

$$F_m = - K_B T N_m \left( ln\left(\frac{Q_{tot,m}}{N_m} + 1\right) \right), \quad \text{for } m = 0,1,2,\ldots, Z_{nuc} \quad (4)$$

where $K_B$ is the Boltzmann constant and the total partition function for the atom/ion $m$, $Q_{tot,m}$, is written in a separable form as

$$Q_{tot,m} = \left(\frac{2\pi m_m K_B T}{h^2}\right)^{\frac{3}{2}} V \cdot Q_{elec,m} \quad \text{for } m = 0,1,2,\ldots, Z_{nuc} \quad (5)$$

where $m_m$ is the mass of the ion of multiplicity $m$, $h$ is Planck's constant and $V$ is the volume of the system. The first factor in the right hand side of (5) is the translational partition function for a classical particle while $Q_{elec,m}$ is the electronic partition function. For the bare aluminum nuclei ($m=13$), the electronic factor is certainly unity. The evaluation of the electronic partition function for the structured particles has been previously studied and discussed in Refs [30,31,33,34]. A possible problem with the evaluation of the electronic partition functions, when using the so-called occupational probabilities, is the cut-off of the ground states in the high density regime causing a nonphysical vanishing of the electronic partition function. This is particularly the case for occupational probabilities that decrease monotonically and indefinitely with density, even for the ground state. To avoid this problem, we use the expression proposed in [34] where

$$Q_{elec}(V,T,\{N\}) = [1 - w_0(V,T,\{N\})] + \sum_{i=0}^{\infty} g_i w_i(V,T,\{N\}) e^{-\varepsilon_i/K_BT} \qquad (6)$$

where $Q_{elec}(V,T,\{N\})$ is the conventional electronic partition function with energy states relative to the electronic ground state and $g_i$ is the statistical weight of the $i$-th level whose energy relative to the ground is $\varepsilon_i$, and $w_i(V,T,\{N\})$ is the state-dependent occupational probability of the $i$-th level. The form of $w_i(V,T,\{N\})$ is usually derived from the plasma microfields and is required to decrease continuously and monotonically with density leading to a physically sensible continuous transition between bound and free states with a natural and smooth truncation of the internal electronic partition function. As it can be seen from Eq. (6) the remedy proposed in Ref. [34] retains at least one state of the ground level in a strongly perturbed system where $w_0(V,T,\{N\})$ goes to zero at low temperatures. The simple expression given by Salzmann [35] for the occupational probabilities is adopted in the present model.

The classical-ideal free energy for free electrons is expressed as

$$F_{e,id}^{cls} = -K_B T N_e \left[ \ln \left( \frac{2V}{N_e} \left( \frac{2\pi m_e K_B T}{h^2} \right)^{3/2} \right) + 1 \right] \quad (7)$$

where $N_e$ is the number of free electrons in the system and $m_e$ is the rest mass of free electrons. Recalling that all interconnected energy levels must essentially be referred to the same reference, hence in the $F_{zero\text{-}of\text{-}energy}$ one has to take into account the sum of the cohesive and ionization energies, i.e.,

$$F_{zero-of-energy} = F_{coh} + \sum_{m=1}^{Z_{nuc}} N_m \sum_{\zeta=1}^{m} I_{\zeta-1} \quad (8)$$

## 4. EXCESS FREE ENERGY COMPONENTS

Coulomb corrections, quantum mechanical effects (partial degeneracy) and other configurational corrections are briefed in Eq. (1) in the term, $F_{corr}$. This term can be expanded into the contributions

$$F_{corr} = F_C + F_{hs} + F_{vdw} + F_{bn} + F_{dgc,\eta} \quad (9)$$

where $F_C$ is the Coulombic excess free energy, which takes into account the interactions among charged particles, $F_{hs}$ is the free energy corrections due to the consideration of particles of finite volumes as a mixture of hard spheres of different diameters, $F_{vdw}$ is the van der Waals' correction, $F_{bn}$ is the correction due to the exclusion of the occupied volume from that accessible to bare nuclei (of vanishing diameters), and $F_{dgc,\eta}$ is the quantum mechanical (degeneracy) correction, with excluded volume taken into account, for free electrons.

With the lack of an exact solution for the many-particle problem of Coulomb-interacting systems, only approximate nonideal plasma models are possible. Accordingly, for the first term in the right hand side of Eq. (9), an approximate cellular model in which the system is regarded as a sum of electrically neutral spherical unit cells, where ions located at the centers of the cells simultaneously neutralize the background electrons, is adopted and the resulting Coulomb correction term to the free energy can therefore be written as

$$F_C \approx -\frac{9}{10}\frac{e^2}{4\pi\varepsilon_0 R_0}\sum_m N_m\, m^2 \qquad (10)$$

where the cell radius $R_0$ is given by $R_0 = (3V/4\pi N_H)^{1/3}$ with $N_H$ representing the number of nuclei in the system and $\varepsilon_0$ is the permittivity of free space. It is useful to recall here that there are many other models of nonideality corrections proposed in the literature such as the Debye-Hückel (DH) model in a canonical ensemble, Single Component Plasma (SCP) model, the Debye-Hückel model in a Grand Canonical Ensemble (GDH), etc. All these models are based on a priori nonideality correction of the free energy function, $F_C$, derived from some model principles. Yet, the cellular model does *not* have the DH limit at low-densities and high-temperatures in contrast to the SCP and GDH models. Compared to the cellular model, the DH, SCP, and GDH models have serious defects such as the existence of more than one root for the solution of the ionization chemical equilibrium problem and the gross violation of the charge conservation law.

The second term in the right hand side of Eq. (9) accounts for the correction to the free energy due to the consideration of the finite size of extended particles (mixture of hard spheres) and is expressed in terms of the packing fraction, $\eta$, as [36,37]

$$F_{hs} = K_B T N \left[ \frac{X\eta}{(1-\eta)^2} + \frac{3Y\eta}{1-\eta} + (X-1)\ln(1-\eta) \right] \quad (11)$$

where $N$ is the number of extended particles in the system and $\eta$, $X$, and $Y$ are given by

$$\eta = \frac{\pi N}{6V} \overline{d^3}, \quad X = \left(\overline{d^2}\right)^3 \Big/ \left(\overline{d^3}\right)^2, \quad Y = \left(\overline{d^2}\right)\left(\overline{d^1}\right) \Big/ \left(\overline{d^3}\right) \quad (12)$$

with

$$\overline{d^k} = \sum_{j=1}^{J} \frac{d_j^k N_j}{N} \quad (13)$$

where $d_j$ and $N_j$ are the diameter and occupation number of the particles of type $j$ in the mixture and $J$ is the number of different extended species in the mixture. For practical implementation of the hard-sphere model described above, one needs to specify the radii of all extended particles employed in the model (neutral atoms and ions of different multiplicities). The use of a compilation of a few thousand of energy levels, in the calculation of internal partition functions of atoms and ions is essentially justified, convenient, and advisable. Conversely, the exhaustive effort of considering such a large number of species of different radii in the calculation of the hard-sphere correction to the free energy lacks similar justification and, for practical reasons, a distinct average hard-sphere radius will be assigned and used for each ion. The hard-sphere radii, $r_m$, for ions of multiplicity $m$ are generally unknown, particularly for the higher charged ions; however different approaches have been proposed and used in the literature to approximate them. For example, Bespalov and Polishchuk [38] used the simple fit

$$r_{m0} = \frac{a_0}{0.086 \left(I_m/eV\right)^{\frac{1}{2}}} \quad (14)$$

where $I_m$ denotes the unperturbed ionization energies and $a_0$ is the Bohr radius. In the relation (14), no effort is devoted to take the dense plasma environment into consideration and therefore $r_{m0}$ represent the radius of the ion $m$ in vacuum. Another approach that takes the dense plasma environment into account is used by Ogata et al [39] where

$$r_{m0} = \frac{r_{00}}{1+0.13\,(I_m/eV)^{\frac{1}{2}}}, \quad \text{and} \quad r_m = r_{m0}\left(1+0.09\,(r_{m0}/D_m)^2\right) \qquad (15)$$

where $r_{00}$ is the radius of neutral atoms in vacuum, $I_m$ is the unperturbed ionization energy of ion $m$ while $D_m$ refers to the screening parameter for the $m$-th ionized atom.

A common proposition in the above and other approaches in the literature is the assumption that the sizes of ions decrease monotonically with increasing ion multiplicity, i.e.,

$$r_{m+1}/r_m = f(m, Z_{nuc}, \ldots) < 1 \qquad (16)$$

Successive use of the relation (16) associates the radii of all ions to a certain reference radius like the atomic radius $r_0$. Fortov e al [40] assumed that the ratio in (16) is approximately constant for all ions. However, for such a highly compressed medium, we believe that the size of the ion will depend on the number of remaining bound electrons to a certain power. Accordingly we suggest the following approximation for the radius of ions of multiplicity $m$ to be considered and investigated with other proposed approximations;

$$r_m = \left((Z_{Nuc} - m)/Z_{Nuc}\right)^n r_0 \qquad (17)$$

where $n$ is a parameter to be investigated.

The choice or determination of the size of neutral aluminum atom is still a problem that needs some investigation. At low densities or vacuum, several estimates for $r_{00}$ can be found in the literature and the choice will, undoubtedly, depend on agreement with available experimental measurements. However, at very high densities, consistency with the cellular model used for Coulomb's correction, as given above, requires that the atomic/ionic size maintains a value not to exceed the Wigner-Seitz (density dependent) cell size. A smoothing formula is proposed and used herein to smooth the transition between the low density constant value $r_{00}$ and the density-dependent Wigner-Seitz radius where

$$r_0(\rho) = \left( r_{WS}^{-3}(\rho) + r_{00}^{-3} \right)^{-\frac{1}{3}} \tag{18}$$

The effect of taking hard-sphere repulsion into consideration is to enhance ionization at very high density in what is usually referred to as pressure ionization. The dependence of $r_0(\rho)$ on the density should be taken into account in the minimization of the free energy and in the derivation of the corrections to thermodynamic functions.

The weak attractive part of the potential between the neutral particles of Al vapor is of importance at relatively low temperatures. The effects of these weak attractive forces are taken into account by means of a van der Waals' excess free energy correction (third term in the right hand side of Eq. (9)). This correction can be roughly expressed as

$$F_{vdw} \approx -\frac{\pi}{3V} N_0^2 \left( \delta_0 \, \sigma_0^3 \right) \tag{19}$$

where $\delta_0$ is the van der Waals' well depth and $\sigma_0$ is the distance at which the interaction energy is zero. Generally, one can consider $\left( \delta_0 \sigma_0^3 \right)$ as one parameter. It has to be noted that the effect of $F_{vdw}$ term on the ionization equilibrium is opposite to the pressure-ionization effect, but it is very weak at temperatures where there is an appreciable amount of ionization. The above treatment of the van der Waals' excess free energy is a very simplified one.

The correction to the free energy due to reduction in the accessible volume for bare nuclei is taken into account in the fourth term in the right hand side of Eq. (9). Bare nuclei can be regarded as point-like particles. The existence of extended particles, like atoms and ions of different multiplicities reduces the volume accessible to bare nuclei, $Al_{13}$. The exclusion of the occupied volume from the volume accessible to bare nuclei gives rise to a correction to the free energy that can be expressed as

$$F_{bn} = -K_B T N_{13} \ln(1-\eta) \tag{20}$$

The correction given by Eq. (20) is equivalent to replacing the volume $V$ in Eq. (5) for $m=13$ by a reduced volume $V^* = V(1-\eta)$ where $\eta$ is the packing parameter given by Eq. (12).

Quantum effects and reduction in the accessible volume for free electrons are taken into account in the last term in the right hand side of Eq. (9). This term can be written as $F_{dgc,\eta} = F_{e,id}^{dgc} - F_{e,id}^{cls}$ where $F_{e,id}^{cls}$ is the ideal classical free energy for free electrons and $F_{e,id}^{dgc}$ is the free energy for electrons with quantum effects (degeneracy) and excluding the volume occupied by the extended particles. Following Ebeling et al [36] and Kahlbaum and Förster [37], $F_{e,id}^{dgc}$ can be written as

$$F_{e,id}^{dgc} = -\left(2 K_B T V^* / \Lambda_e^3\right) \theta_{3/2}(\mu_{e,id}^* / K_B T) + N_e \mu_{e,id}^* \tag{21}$$

where $\Lambda_e = h/\sqrt{2\pi m_e K_B T}$ is the average thermal wave length of the electrons,

$V^* = V(1-\eta)$ is the reduced volume, $\theta_\nu$ is the complete Fermi-Dirac integral, and $\mu_{e,id}^* = \partial F_{e,id}^{dgc}/\partial N_e$ is the chemical potential of the ideal Fermi electron gas related to the number of free electrons in the system by

$$N_e = \left(2V^*/\Lambda_e^3\right)\theta_{1/2}(\mu_{e,id}^*/K_B T) \tag{22}$$

It has to be noted that densities per unit volume are taken always according to the total volume $V$.

At equilibrium, occupation numbers of all species can be found by minimizing the free energy function in the ionization/recombination processes subjected to the constraints of electroneutrality and conservation of nuclei. This approach of free energy minimization is known to assure thermodynamic consistency among the population numbers and different thermodynamic functions derived from the same free energy function. The computational scheme used to solve the problem, for a constant temperature, $T$, and a specified mass density, $\rho$, and the high pressure corrections to thermodynamic properties are explained elsewhere [30,31,41,42].

## 5. RESULTS AND DISCUSSIONS

Estimates of the critical point of Al are obtained through inspection of the equation-of-state isotherms and numerically identifying the point of inflection, characterized by

$$(\partial P/\partial V)_T = \left(\partial^2 P/\partial V^2\right)_T = 0 \tag{23}$$

For isotherms below the critical isotherm a Maxwell construction is used to remove the developed van der Waals loops. This procedure is shown in Fig. 1 in which a set of pressure isotherms of aluminum, including the critical isotherm, is presented with the use of Maxwell construction for isotherms with $T<T_C$.

There are effectively three parameters in the present model; the radius of the neutral atom in vacuum ($r_{00}$), the parameter $\delta_0\sigma_0^3$ in the van der Waals correction and the power $n$ to which the radius of an ion depends on the number of remaining bound electrons. Computational results showed that the low temperature results around the critical point are sensitive to $r_{00}$ and $\delta_0\sigma_0^3$ more than to the value of $n$ while results in the high density and relatively high temperature regime are more sensitive to $n$ than to $r_{00}$ or $\delta_0\sigma_0^3$. Accordingly, the values of $r_{00}$ and $\delta_0\sigma_0^3$, used in the present computations, are chosen based on agreement with available values of the density of the liquid-phase at temperatures below the critical point.

The isotherms presented in Fig. 1 are calculated using $n=0.5$, however, several similar calculations are performed for values of the parameter $n$ between $n=1/3$ and $n=1.0$ with the resulting values of the critical parameters summarized in Table 2. The value of $n=1/3$ corresponds to the assumption that the volume of an ion is directly proportional to the number of bound electrons remaining in the ion. As it can be seen from the table, the critical parameters of Al are not crucially sensitive to $n$ and can be roughly taken to be the average of these values as shown in the last row in the table. It has to be noted that the present predicted values of the critical parameters lie within the range of values in the literature as summarized in Table 1. Other formulae for the radii of ions of different multiplicities similar to those given by Eq. (14) and Eq. (15) have been examined; however, they produced unphysical intersecting isotherms

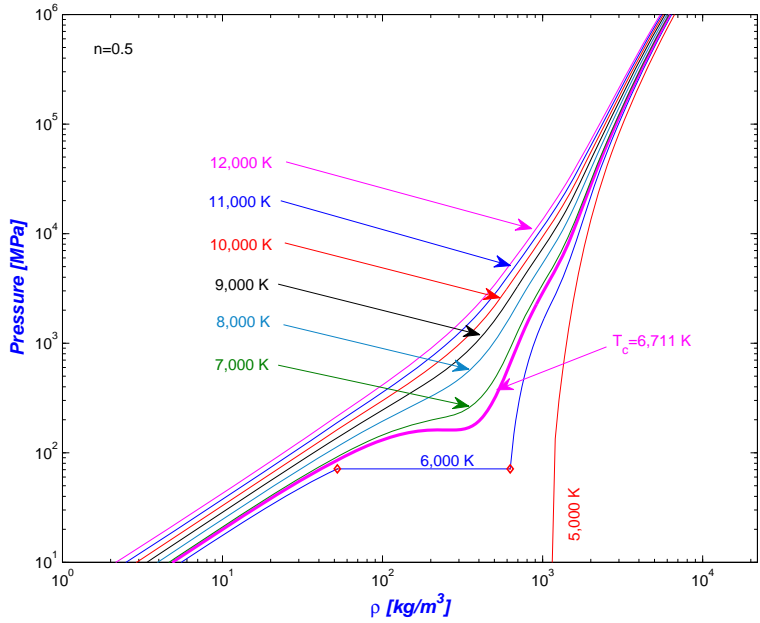

Figure 1 Aluminum isotherms around the critical point with Maxwell construction showing phase transition for temperatures lower than $T_c$

Table 2. Dependence of the critical parameters of Al on the parameter $n$ in Eq. (17). Present results are calculated using $r_{00}=2.11$ Å

| $n$ | $T_c$[K] | $P_c$ [MPa] | $\rho_c$ [kg m$^{-3}$] |
|---|---|---|---|
| 1/3 | 6723 | 159.3 | 246.8 |
| 0.5 | 6711 | 162.0 | 230.1 |
| 0.6 | 6705 | 160.3 | 229.6 |
| 0.7 | 6698 | 160.5 | 230.4 |
| 0.8 | 6691 | 160.6 | 230.2 |
| 0.9 | 6685 | 160.2 | 230.4 |
| 1.0 | 6678 | 159.6 | 230.2 |
| Average Values | 6699 | 160.4 | 232.5 |

Isotherms of the total pressure and of the specific internal energy of Al fluid are shown in Fig. 2 and Fig. 3, respectively. These two *explanatory* figures are generated, ignoring the photon gas, using *n=0.5*. The pressure appears to behave ideally at low

Page 16

densities and/or high temperatures as expected. However, deviations from this ideal behavior develops for high densities and relatively low temperatures and become most prominent with the critical isotherm (lowest temperature shown). Similar observations can be noted for the isotherms of the specific internal energy shown in Fig. 3.

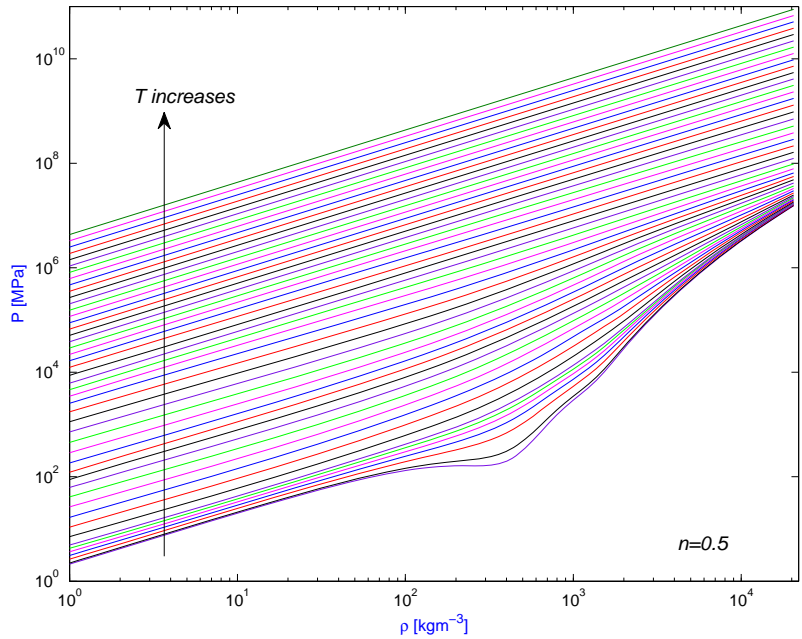

Figure 2. Pressure isotherms of Al fluid as a function of density; the lowest isotherm is the critical isotherm ($T_c$ = 6,711 K) for *n=0.5*, then isotherms from 7,000 to 12,000 K with an increment of 1,000 K followed by isotherms from 15,000 to $10^9$ K which are equally spaced on the logarithmic scale with $\Delta log_{10}(T)=0.1206$.

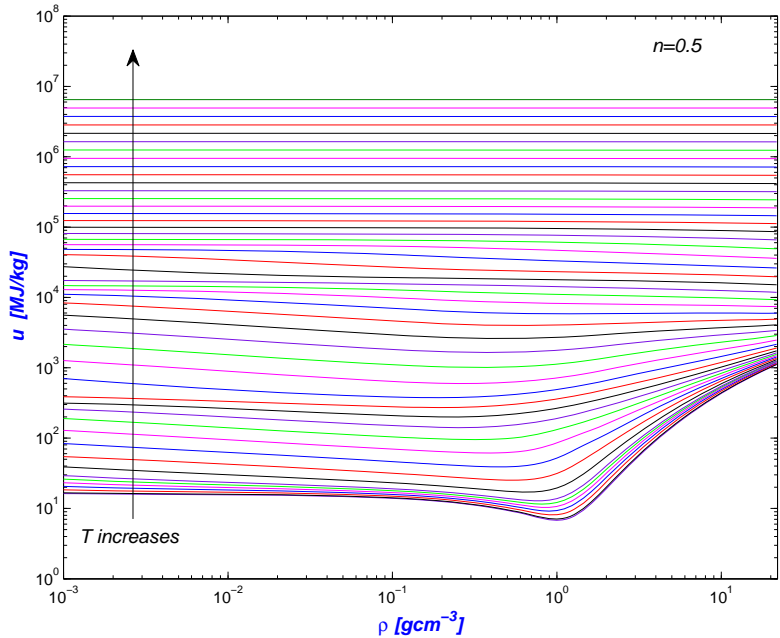

Figure 3. Isotherms of the specific internal energy of Al fluid as a function of density; the lowest isotherm is the critical isotherm ($T_c$ = 6,711 K) for *n=0.5*, then isotherms from 7,000 to 12,000 K with an increment of 1,000 K followed by isotherms from 15,000 to $10^9$ K which are equally spaced on the logarithmic scale with *$\Delta log_{10}$ (T)=0.1206*.

Figure 4 part(a) shows a comparison between the principal shock Hugoniot (pressure *P* vs compression ratio $\rho/\rho_0$) predicted from the present equation-of-state model, considering four different values of the parameter *n*, and relevant experimental data available in the literature [43-45]; part (b) shows the corresponding curves of the temperature *T* vs the compression ratio $\rho/\rho_0$, while part(c) shows the corresponding curves of the average ionization state <Z> vs the compression ratio $\rho/\rho_0$. For very strong shock waves where the pressure and temperature behind the front are much greater than the initial pressure and initial temperature, ionization is complete and the plasma approaches an ideal gas of bare nuclei and free electrons with the compression ratio across the shock approaching its ideal gas asymptote $\rho/\rho_0=4$. At intermediate to high shock pressures and temperatures, the material becomes partially to fully-ionized and the equation-of-state depends to a large extent on the electronic structure. A maximum compression, beyond the ideal-gas asymptote, $\rho/\rho_0=4$, is attainable in this region due to the absorption of energy in the excitation and ionization processes. Nevertheless, this region in the aluminum Hugoniot embodies two kinks or shoulders resulting from the competition between the absorption of internal energy for ionizing the shells (enhances compression) and the increase in pressure due to release of more free electrons which opposes compression. These two shoulders in the Hugoniot curve of Al have been recognized and reported in more than one place in the literature and are due to the successive ionization of the $L$ and $K$ shells (see for example, [40, 46-51]). Even though, predictions from the present model show a third kink at lower temperature/pressure (weaker shock) which can be attributed to successive ionization of the three electrons in the $M$ shell. It is a well-known fact that filled shells generally require high ionization energies and jumps always exit between the ionization potentials of the last electron in a shell and the first electron of the following shell [52]. For example, there is a jump between the ionization potential of the 3$^{rd}$ ionization state, last in the $M$ shell (28.446 eV for a free Al atom), and the 4$^{th}$ ionization state, first in the $L$ shell (119.98 eV for a free Al atom). Similarly there is a larger jump between the last $L$ electron (441.97 eV for a free Al atom) and the first $K$ electron (2085.8 eV for a free Al atom). Because of these jumps in the ionization potentials, different levels of shock strengths would be required for the removal of the electrons of each shell as it appears clearly in Figs. 4(a) and 4(b). It is a well-known fact that ionization energies are strongly perturbed and are generally lowered in a dense plasma system which agrees with Fig 4 (b) where the locations of the three shoulders are lower than the ionization energy span for the corresponding shell in a free atom (5.99-28.45 eV for the M-shell, 119.99-441.99 eV for the L-shell and 2085.98-2304.14 eV for the K-shell). It has to be mentioned that spectroscopic data (excitation levels, statistical weight, ionization energies, etc.) used in the present computations are taken from NIST compilations [53]. Figure 4(c) shows clearly without reservation that the kinks in the shock Hugoniot including the third kink are due to the successive ionization of electronic shells. This part of the figure interestingly shows that the start and end of the kink follow the beginning and completeness of the ionization of the corresponding shell.

Although this interpretation of the appearance of this third kink in the Hugoniot of the present results appears to be rational and justified, the absence of this third kink in the Hugoniot curves obtained from other well-received theoretical models represents a puzzle that needs to be solved. Figure 5 shows a comparison between the Hugoniot curves calculated from the present work and those calculated by Pain [49] using Thomas-Fermi model, Equation of State with Orbital Description of Electrons (ESODE) with two treatments for the thermal electronic contribution to the equation of state: average atom in a spherical cell (ASC) and average atom in a jellium of charges (AJC) in addition to the results of Piron and Blenski [50] using Variational-Average-Atom-in-Quantum-Plasma (VAAQP). Figure 6 also shows a comparison between the Hugoniot curves from the present work and those calculated by Kadatskiy and Khishchenko [51] using Hartree-Fock-Slater (HFS) model where the influence of the thermal motion and interaction of ions are taken into account in the framework of three models: the ideal gas (IG), the one-component plasma (OCP) and the charged hard sphere (CHS). As can be seen from these figures (Fig. 5 and Fig. 6) the third kink appears only in the results of the present model.

Although experimental measurements reserve the final word in this situation, the fact that most of the available experimental data are either reported with large experimental uncertainity (as can be seen, for example, in Fig. 4(a)) or reported without any experimental uncertainity weakens any conclusion, derived based on comparison with currently available experimental data, towards understanding the origin of this third kink in our results and its absence in the results of other models. Even though one can still count on theoretical reasononing and physical insight. In this regard it may be useful to refer to a recent work by Driver and Militzer [54] where the Hugoniot curve of nitrogen calculated using first principle Path Integral MonteCarlo (PIMC) simulations shows a third shoulder as appears in Fig. 13 in [54] which has been attributed to molecular dissociation. This finding backs our finding of the third kink or shoulder in the Al Hugoniot (attributed to ionization of the M shell). The effect of the ionization of the M shell electrons (three ionization processes) is similar to if not more prominent than the effect of dissociation of molecules in producing a shoulder in the Hugoniot curve. Perhaps first principle PIMC simulation can help in solving this puzzle of the third kink.

A reconciliation with the results of [49,51] may be sought through interpreting the absence of the third kink in the results of these models as a result of considering the zero-degree isotherm for the solid phase while the present model consideres only alumium fluid or more precisely supercritical aluminum gas. Accordingly, if a phase change from supercritical aluminum gas to solid aluminum occurs, one has to consider/add the lattice or zero-degree pressure and internal energy to those of the high temperature fluid aluminum. Depending on the relative magnitude of the pressure change through the third kink to the corresponding lattice pressure the third kink may or may not appear in models adding the zero-degree pressure and internal energy beforehand.

To roughly estimate the effect of considering zero-degree isotherm on Al Hugoniot curve, we add the zero-degree pressure and internal energy (which are functions of the density) to those calculated from the present model and recalculate the shock Hugoniot of Al. Figure 7 shows the results of such computations adding the zero-degree pressure and internl energy according to Naumann [55]. As it can be seen from the figure the third kink disappears with all values of the parameter $n$ except $n=1/3$ where it fades out to some extent. It has to be noted that the zero-degree pressure and internal energy should vanish in the gas phase. Accordingly, if a phase change from the supercritical gas to the solid phase occurs in the region of the third kink, it would not appear. However, if no phase change occurs in this region, the third kink should appear and the inclusion of the zero-degree pressure and internal energy in other models lacks its justification in this case. Again, the final word in this problem is due to accurate experimental measurements in this region and if one considers the collection of experimental measurements presented in Fig. 2 of Kadatskiy and Khishchenko [51] irrespective of the missing experimental uncertainty, the first scenario of implying a phase change would be the likely one. Even though the fact that the results of Piron and Blenski [50] showed in Fig. 5 are obtained from a plasma equation of state using the VAAQP electronic contribution and an OCP ion contribution with no zero-degree isotherm correction, as clarified by Piron [56], weakens the above reconciliation and leaves the question unresolved. In all cases, important and interesting physics exist and will be explored in this region.

Despite of the sensitivity of the present predictions of Al Hugoniot to the parameter *n* or equivalently to the size of ions as shown in Figs. 4-6, all curves for the examined values of *n* show an overall good agreement with the available relevant experimental data. In addition, and except for the appearance of a third lower temperature/pressure kink in the Hugoniots of the present model which has been discussed above, the behavior and order of magnitude of the results are in fair agreement with other theoretical predictions in the literature [40, 46-51].

The effect of including the photon gas on the calculation of the shock Hugoniot of aluminum is shown in Fig. 8 where the Hugoniot calculated considering the photon gas approaches the well-known theoretical limiting value of $\rho/\rho_0=7$ characteristic to the photon gas.

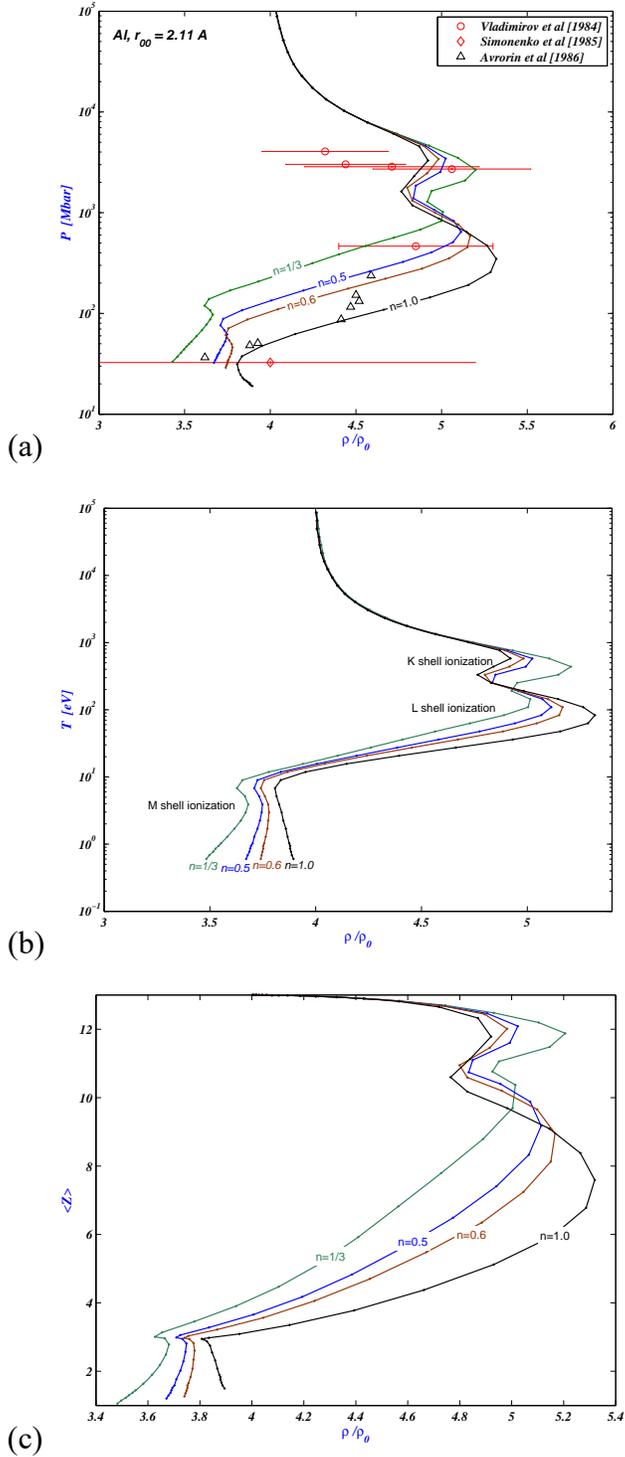

Figure 4. (a) Comparison of the calculated shock Hugoniots of Al (pressure $P$ vs compression ratio $\rho/\rho_0$) with experimental data, (b) Temperature through the Hugoniot as a function of the compression ratio $\rho/\rho_0$, and (c) Average ionization state $<Z>$ through the Hugoniot as a function of the compression ratio $\rho/\rho_0$.

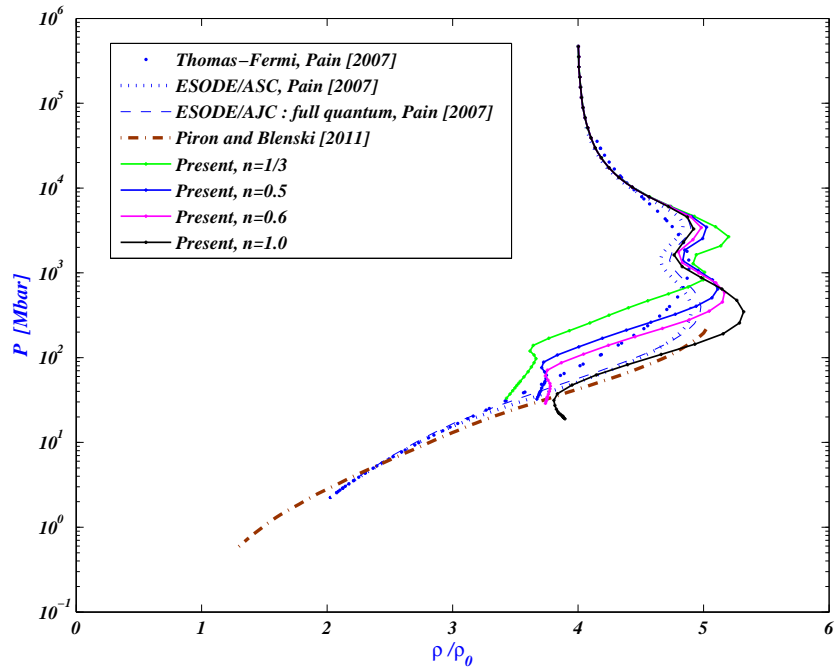

Figure 5. Comparison of the calculated shock Hugoniots, as a function of the compression ratio $\rho/\rho_0$ of Al, with predictions of Pain [49] and Piron and Blenski [50].

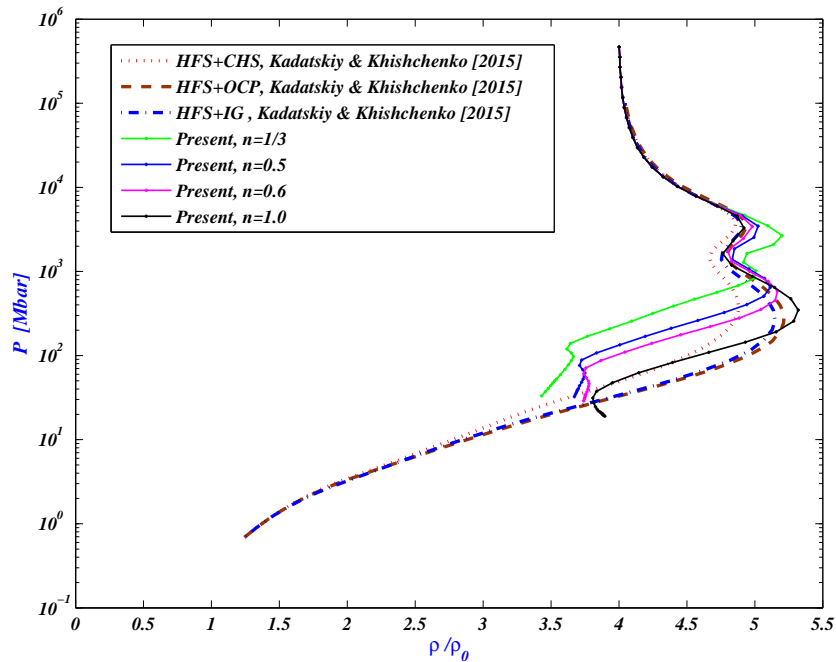

Figure 6. Comparison of the calculated shock Hugoniots, as a function of the compression ratio $\rho/\rho_0$ of Al, with predictions of Pain [51].

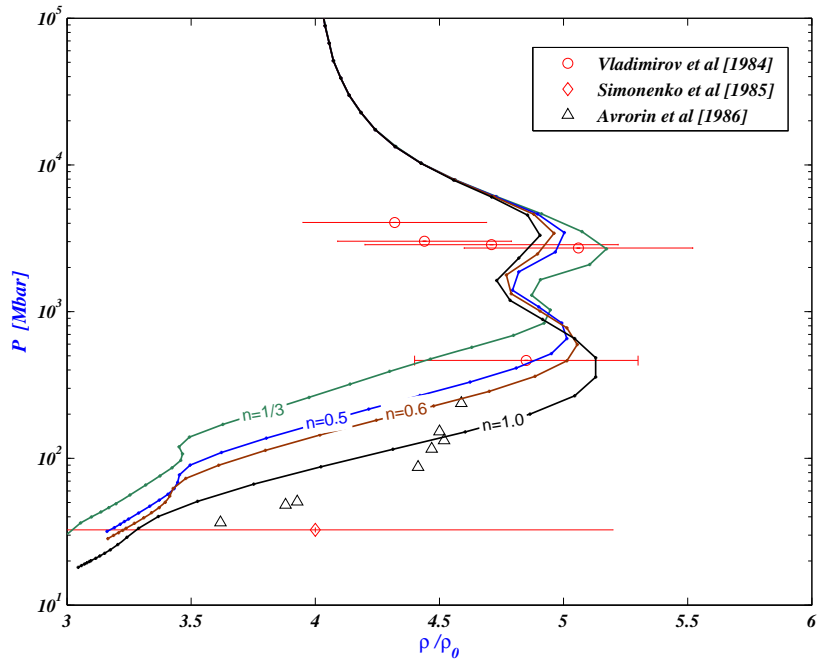

Figure 7. Comparison of the calculated shock Hugoniots, as a function of the compression ratio $\rho/\rho_0$ of Al, with the addition of the zero-degree pressure and internal energy according to Naumann [55].

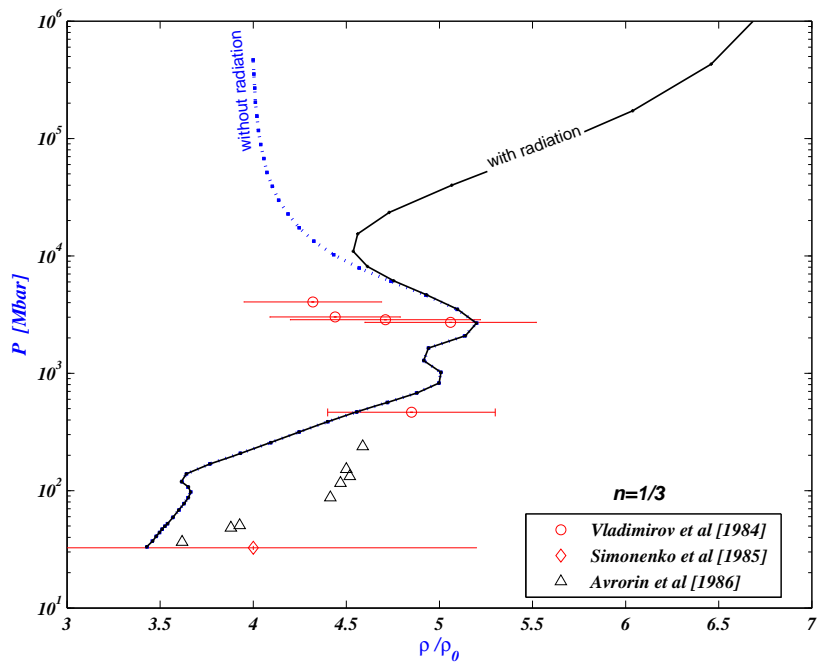

Figure 8. Effect of including the photon gas on the calculation of the Hugoniot curve for Al ($n=1/3$).

## 6. CONCLUSIONS

Preliminary estimates of the critical properties ($T_c \sim 6699$ K, $P_c \sim 160.4$ MPa, $\rho_C \sim 232.5$ kg/m3) of pure aluminum fluid are determined from a chemical model for the equation-of-state of hot dense partially ionized fluid. The model is used to generate the thermodynamic properties of aluminum and to calculate the principal shock Hugoniot curve. The calculated Hugoniot curves are in fair agreement with available experimental measurements and show the expected limiting behavior of an ideal gas $\rho/\rho_0)_{max} = 4$ and of a photon gas $\rho/\rho_0)_{max} = 7$ for very strong shocks (extremely high temperatures and high pressures). A distinctive feature of our calculated Hugoniots is the appearance of a third kink at low temperature/pressure due to successive ionization of the electrons in the $M$ shell which has been discussed and explained. Present predictions of the critical properties of Al and the maximum compression ratio are in line with the range of values reported in the literature.

## ACKNOWLEGMENTS

The author acknowledges useful and valuable comments and suggestions received from the anonymous reviewer. The author would like also to thank Jean-Christophe Pain, CEA/ France, Robin Piron, CEA/France and M. A. Kadatskiy, JIHT of the Russian Academy of Science, Russia, for providing the necessary information and data to reproduce the Hugoniot curves predicted by their theoretical models and for fruitful discussions. I would like to extend my appreciation to R. Redmer and B. Whitte, University of Rostock, Germany, for their useful feedback and suggestions.